\documentclass[journal=jctc,manuscript=article,layout=twocolumn]{achemso}

\usepackage[version=3]{mhchem} 
\usepackage{lscape}

\usepackage{amssymb}
\usepackage{amsmath}

\usepackage{xcolor}

\SectionNumbersOn

\author{Xiangyue Liu}
\affiliation{Fritz-Haber-Institut der Max-Planck-Gesellschaft, Faradayweg 4-6, D-14195 Berlin, Germany}
\author{Laura McKemmish}
\affiliation{School of Chemistry, UNSW Sydney, Sydney, NSW 2052, Australia}
\author{Jesús Pérez-Ríos}
\email{jperezri@fhi-berlin.mpg.de}
\affiliation{Fritz-Haber-Institut der Max-Planck-Gesellschaft, Faradayweg 4-6, D-14195 Berlin, Germany}
\title{Is CCSD(T) a proper standard for dipole moment calculations? An analysis considering diverse diatomic species. }

\abbreviations{}

\begin{document}



\begin{abstract}
Coupled cluster with single, double, and perturbative triple excitations [CCSD(T)] has been extensively employed as the reference method in benchmarking different quantum chemistry methods. In this work, we test the accuracy of CCSD(T) calculating ground state electric dipole moments at the extrapolated complete basis set (CBS) limit. The calculated dipole moments have been compared to an experimental dataset consisted of diatomic molecules with various kinds of bond natures and spin configurations. As a result, to reach a satisfactory agreement with experimental dipole moments, core-correlations should be included for some molecules. However, even when core-correlations are included, the predicted dipole moment deviates considerably from the experimental values for molecules involving transition metal atoms.  
\end{abstract}

\small

\section{Introduction}

Coupled cluster with single, double, and perturbative triple excitations, CCSD(T), is one of the most popular methods serving as a benchmark reference in developing electronic structure theory methods such as density functional theory, DFT. It is size-consistent, and as a member of the coupled cluster family, it is systematically improvable. When utilized in combination with certain corrections, it is considered to approach sub-chemical accuracy in properties such as bond energies \cite{fang2017prediction} at the complete basis set (CBS) limit \cite{augccdunning1989gaussian}. 

In the DFT front, benchmarking studies most commonly focus on energetic properties, but recently there has been increased interest in understanding performance in non-energetic properties most notably the electric dipole moment\cite{dipoleDFThait2018accurate,dipoleDFThait2018communication,dipoleDFTjohnson2019effect,dipoleDFTgrotjahn2020evaluation,zapata2020computation}. These benchmark studies generally benchmark by comparing approximate DFT results against CCSD(T) calculations, which are viewed as the ``gold standard'' and assumed as accurate. However, this assumption has not been tested (and may not be accurate) for a diverse range of systems, particularly those containing transition metals or for van der Waals complexes. 

The primary aim of this paper is to quantify the accuracy of the CCSD(T)/extrapolated-CBS approximation against our recently collated set of 135 experimental dipole moments \cite{dipolePCCP2020} for diatomic molecules. These diatomics - containing both main-group and transition metal elements - serve as model systems\cite{furche2006performance} for diverse set of species with different bond natures and spin-configuration; bi-alkali molecules are the simplest models of van der Waals interaction \cite{zhao2006comparative}, transition metal diatomics provide test cases for multi-reference systems \cite{jensen2007performance,yanagisawa2000investigation,buhl2006geometries,barden2000homonuclear,furche2006performance,gutsev2003chemical, schultz2005databases, sorkin2009energies, cramer2009density,16TeLoMc}, while performance on highly ionic systems can be tested using alkali metal halides. The simplicity of diatomic molecules enables thorough evaluation of the importance of high-order corrections - such as core-valence correlation - that would be infeasible in larger systems.

The paper is structured as follows. We describe the computational approach and the dataset in Sec.\ref{sec:method} and \ref{sec:dataset}, respectively. In Sec.\ref{sec:frozen-core}, we investigate the performance of standard frozen-core CCSD(T)/extrapolated-CBS calculations in predicting diatomic equilibrium distances and dipole moments. The role of core-correlation is quantified and discussed in  Sec.~\ref{sec:core0}. Basis set effects are discussed in Sec. \ref{sec:cbs} and Sec. \ref{sec:diffuse}. Finally, we investigate whether using experimental rather than computational equilibrium distances yields substantial improvement in the predicted dipole moments in Sec.\ref{sec:opt_vs_exp_Re}. 

\section{Methodology}
\subsection{\label{sec:method}Computational approach}
The calculations are performed using the Molpro package~\cite{molpro,werner2012molpro}. The equilibrium geometries of the molecules are optimized at the spin-unrestricted CCSD(T) level with force tolerance of $3 \cdot 10^{-4}$ a.u. and maximum energy change of $10^{-6}$ Hartree. The dipole moments are obtained from four-point central finite differences of the total energy upon finite electric field applied as perturbation with a strength of $10^{-3}$ a.u. and $2\cdot 10^{-3}$ a.u., respectively. In this work, only the magnitude of the dipole moments is discussed.

Calculations for all molecules were performed using the segmented def2- series basis sets developed by Ahlrichs \textit{et. al}, including def2-TZVPP and def2-QZVPP \cite{def2weigend2003gaussian,def2weigend2005balanced}, using effective core potentials for elements with $Z > 36$. For molecules containing only elements with $Z\le 36$ (Kr), we additionally consider results from the larger aug-cc-pVQZ  and aug-cc-pV5Z augmented Dunning's correlation-consistent polarized basis sets \cite{augccbalabanov2005systematically,augccdunning1989gaussian,augcckendall1992electron,augccwilson1999gaussian,augccwoon1993gaussian,augccwoon1995gaussian,augccfeller1996role}. 

For the Dunning basis sets, the CBS limits can be predicted using the standard two-point extrapolation scheme as 

\begin{equation}
\label{eq:CBS}
    \text{Predicted CBS}(n_1/n_2) = \frac{n_1^3 d_1 - n_2^3 d_2}{n_1^3 - n_2 ^3},
\end{equation}

\noindent
where $d_i$ is a molecular property (equilibrium internuclear distance $R_e$ and dipole moment, $\mu$, in our work) while $n_1$ and $n_2$ are 4 and 5 for the aug-cc-p(C)VQZ and aug-cc-p(C)V5Z basis sets, respectively. Although the def2- basis sets were not designed explicitly for extrapolation, CBS extrapolation techniques have been shown to deliver very accurate results for some molecules with light elements \cite{neese2011revisiting}. Therefore, we employ the same extrapolation scheme as in Eq.~\ref{eq:CBS}, with $n_1$ and $n_2$ being 3 and 4 for the def2-TZVPP and def2-QZVPP basis sets, respectively.

\subsection{\label{sec:dataset}Dataset}

\begin{table*}
  \caption{Molecules in the dataset classified by groups of their constituent elements.}
  \label{table:molecule_class}
  \begin{tabular}{p{7cm}p{8cm}}
\hline
Types of molecules  &  Molecules\\ 
\hline
alkali metal - alkali metal & LiK, LiNa, LiRb, NaCs, NaK, NaRb \\
halogen - alkali metal & CsBr, CsCl, CsF, CsI, KBr, KCl, KF, LiBr, LiCl, LiF, LiI, NaBr, NaCl, NaF, NaI, RbBr, RbCl, RbF, RbI \\
halogen - alkaline earth & BaF, CaF, CaI, SrF \\
halogen - halogen & ClF, IBr, ICl, IF \\
metal/metalloid - halogen & AlF, BF, GaBr, GaF, InCl, InF, TlBr, TlCl, TlF, TlI \\
metal/metalloid - metal/metalloid & GeTe, PbTe, SnTe \\
nonmetal - alkali metal & LiH, LiO, NaH \\
nonmetal - alkaline earth & BaO, BaS, CaD, CaH, MgD, MgO, SrO \\
nonmetal - halogen & BrO, CF, ClD, ClH, ClO, DBr, DF, HBr, HF, HI, ID, OF, SF, SeF \\
nonmetal - metal/metalloid & BH, GeO, GeS, GeSe, PbO, PbS, PbSe, SiO, SiS, SiSe, SnO, SnS, SnSe \\
nonmetal - nonmetal & CH, CN, CO, CS, CSe, HD, NH, NO, NS, OD, OH, PN, PO, SD, SH, SO, SeD, SeH \\
transition metal - halogen & AgBr, AgF, AgI, AuF, CuF, HfF, IrF, YF \\
transition metal - nonmetal & AgH, AuO, AuS, CrO, CuO, CuS, FeO, HfO, IrC, LaO, MoN, NbN, NiH, PtN, PtS, ReN, RhN, ScO, ScS, TiN, TiO, VN, VO, VS, WN, YO, ZrO \\
\hline   

  \end{tabular}
\end{table*}

\begin{figure}[h!]
    \centering    
    \includegraphics[width=0.5\textwidth]{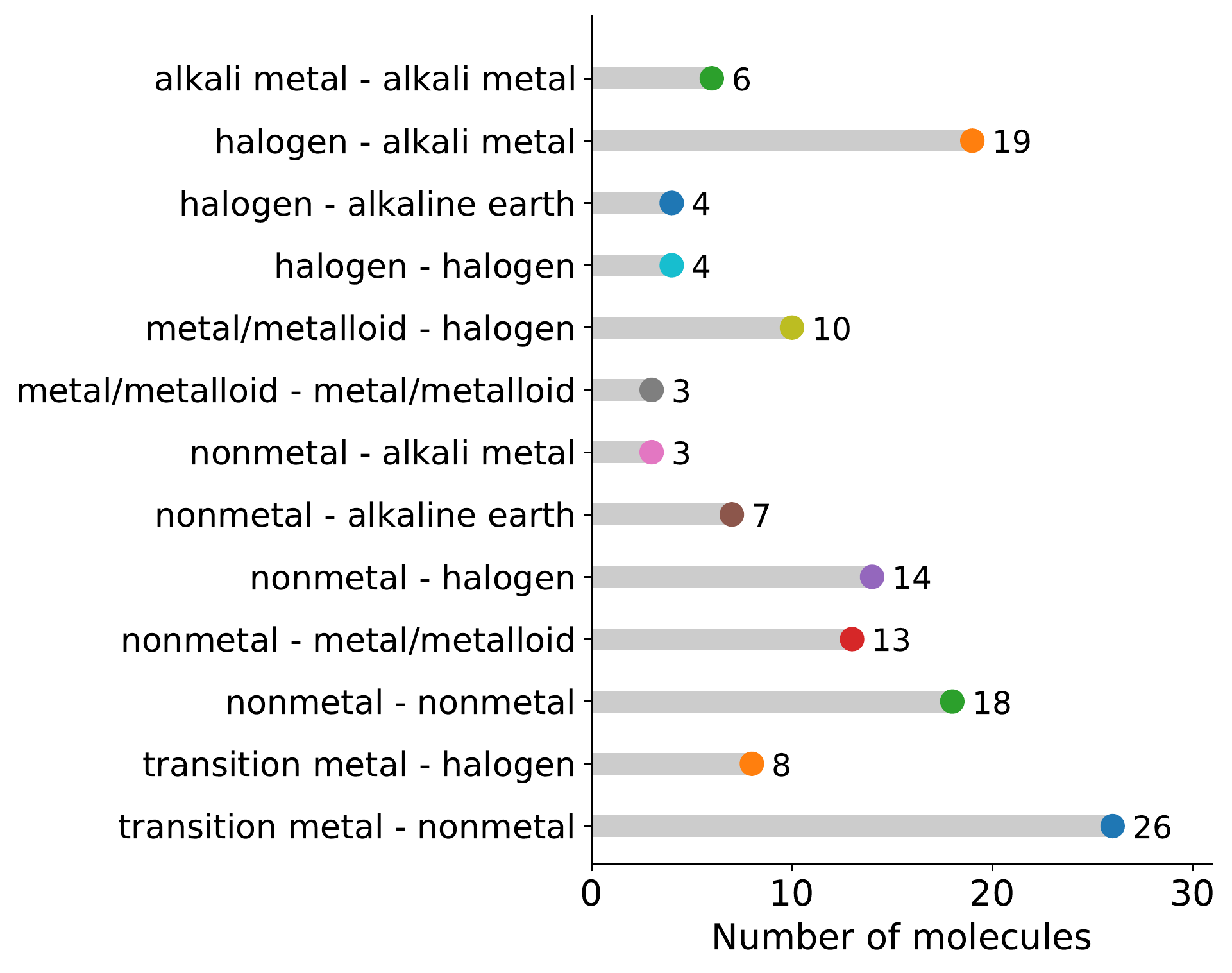}
    \caption{Number of molecules in the present dataset, classified by groups of their constituent elements.}
	\label{fig:molecule_class}
\end{figure}

The dataset consists of 135 heteronuclear diatomic molecules taken from Ref.\cite{dipolePCCP2020}, with experimental ground-state $R_e$ and dipole moments. The molecules in the dataset can be classified by types of their constituent elements, as shown in Fig.\ref{fig:molecule_class} and Table \ref{table:molecule_class}. The dataset contains various main-group metal and non-metal compounds of different types of bonds. In particular, several van der Waals molecules, e.g., bi-alkali, are included in this dataset. Similarly, the dataset contains 34 transition metal compounds, including f-block elements.

\section{Results and discussions}

In this section, we present and discuss the theoretical predictions of the equilibrium distance, $R_e$, and dipole moments compared to their experimental values for the molecules in the dataset. We quantify differences between experimental and our calculated values using residual (calculated - experimental) and root-mean-squared error (RMSE). 

For each computational method investigated, we present our results in two formats. First, we present residual errors of all molecules (as a function of experimental equilibrium distance or dipole moment), labeling molecules with significant errors. Second, we present RMSE grouped by molecular class as designated in Table \ref{table:molecule_class}. Together, these figures enable us to understand the expected accuracy of different types of CCSD(T)/extrapolated CBS calculations within different molecule classes.

\subsection{\label{sec:frozen-core}Frozen-core approach using different basis sets}

\begin{figure*}[h!]
    \centering    
    \includegraphics[width=1\textwidth]{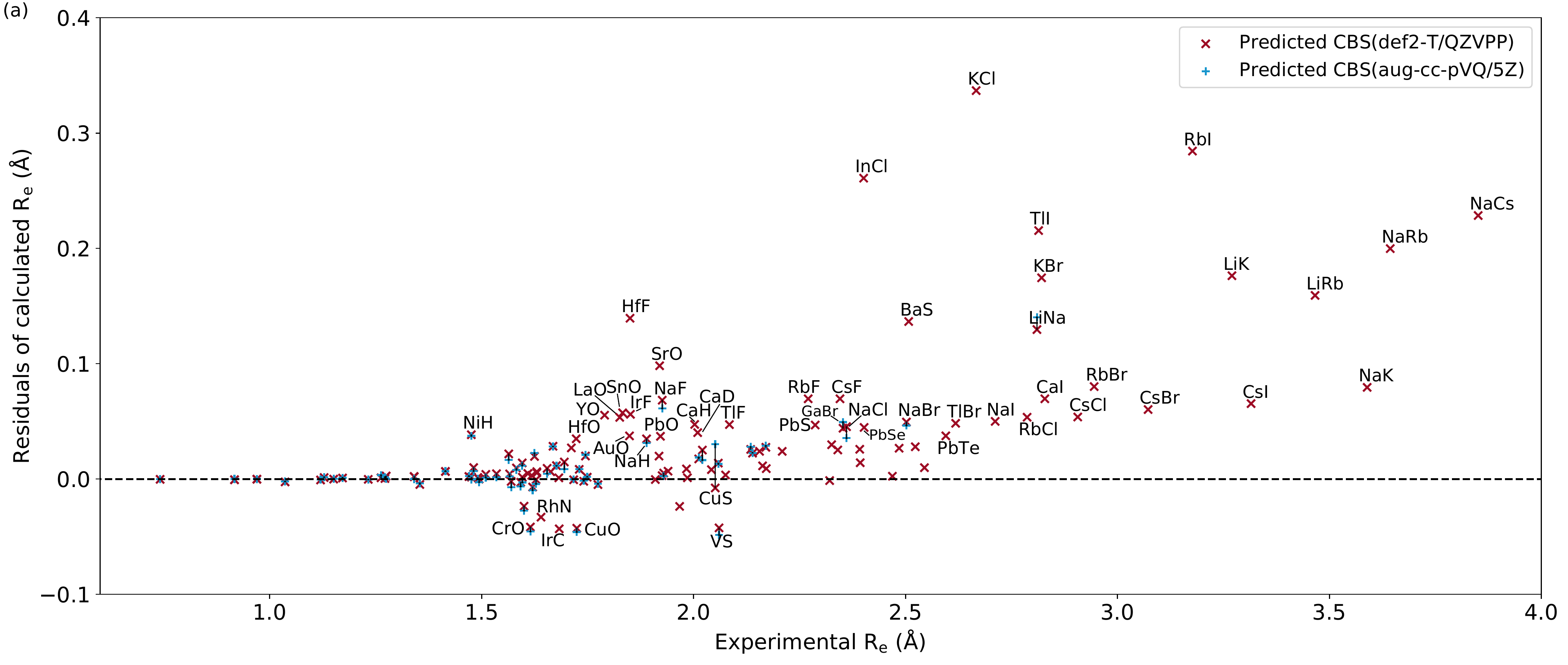}\\
    \includegraphics[width=1\textwidth]{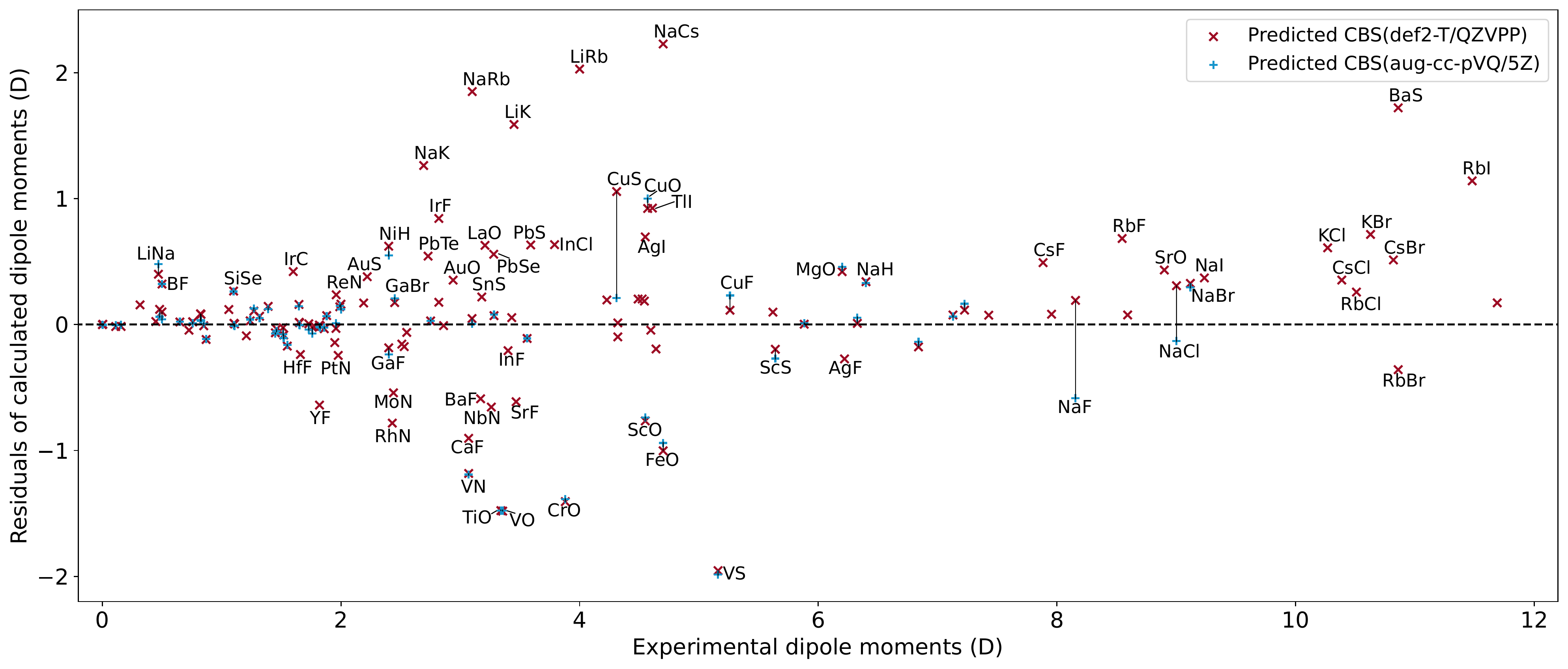}
    \caption{Residuals of calculated (a) $R_e$ and (b) dipole moments compared to experimental values under frozen-core approximation. Shown are the predicted CBS of $R_e$ or dipole moments from def2-T/QZVPP (135 molecules) and aug-cc-pVQ/5Z (67 molecules) basis. Vertical lines joining dipole moments by two CBS' are added to guide the eye.}
	\label{fig:residual_Re_dipole_frozen_core_CBS_def2tqz_av45z}
\end{figure*}

\begin{figure}[h!]
    \centering    
    \includegraphics[width=0.5\textwidth]{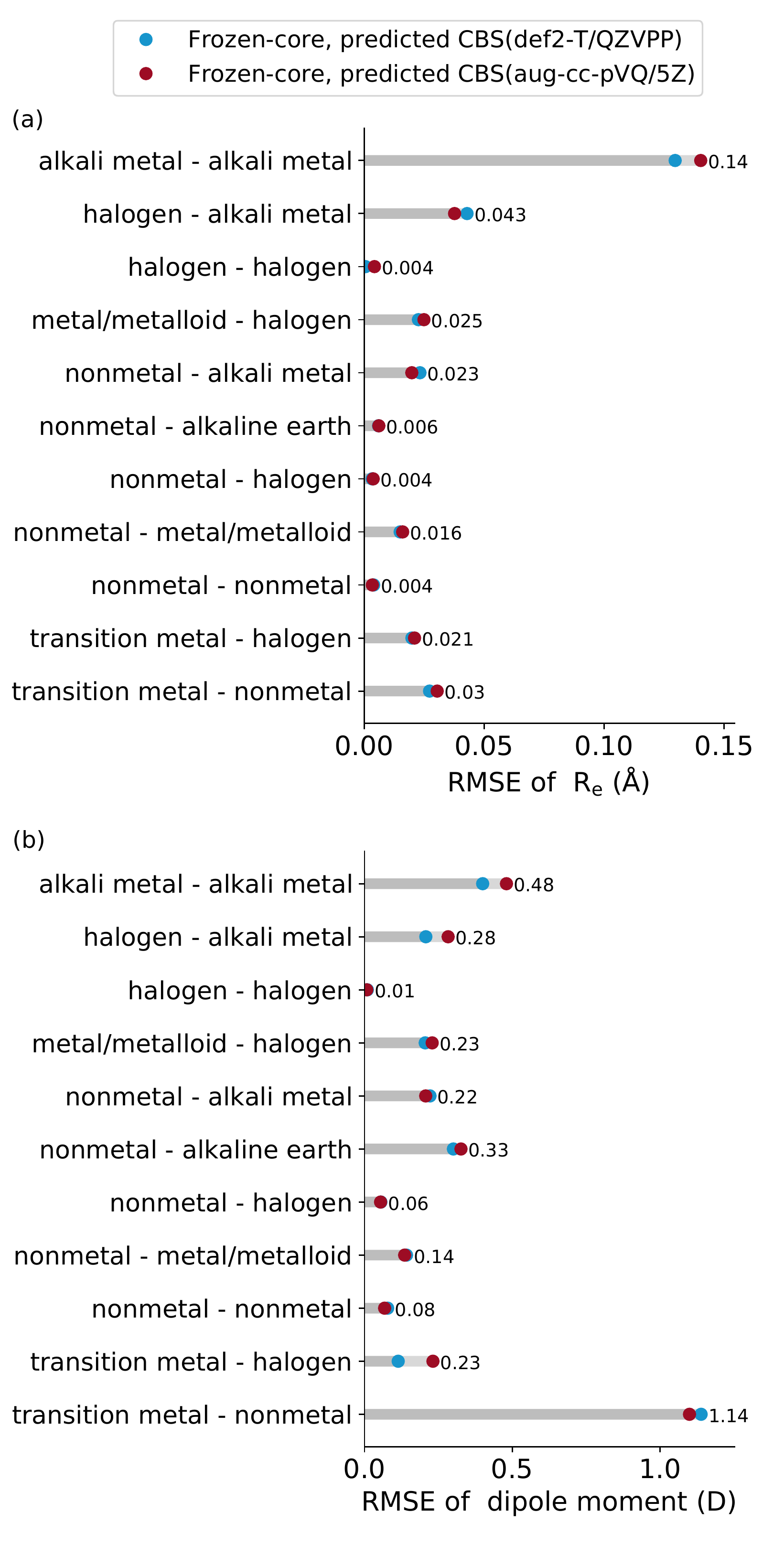}
    \caption{RMSE of calculated (a) $R_e$ and (b) dipole moments predicted obtained from frozen-core CCSD(T)/CBS(def2-T/QZVPP) and predicted CBS(aug-cc-pVQ/5Z) for 67 molecules consisted of elements with $Z<36$, classified by the types of molecules.}
	\label{fig:RMSE_class_frozencore_CBS_def2tqz_av45z_dipole_Re}
\end{figure}

The default standard approach with CCSD(T) calculations is usually frozen-core calculations. Our frozen-core calculations are presented in Fig.~\ref{fig:residual_Re_dipole_frozen_core_CBS_def2tqz_av45z} and \ref{fig:RMSE_class_frozencore_CBS_def2tqz_av45z_dipole_Re}. 

The RMSE for equilibrium distance for our set of 135 diatomics is 0.070 \AA{} when calculated using frozen-core CCSD(T)/CBS(def2-T/QZVPP). Usually, errors in equilibrium distance are less than 0.01 \AA{} (e.g., non-metal diatomics in def2- results), but outliers such as KCl, NaCs have errors above 0.2 \AA{}, certainly too high to be a trusted ``gold standard'' calculation. Indeed, for some molecules, a more accurate result can be obtained by using machine learning techniques using only atomic properties of the atoms constituting a molecule~\cite{Liu2021}. Diatomics with longer bond lengths generally have larger errors. Based on the identified outliers in Fig.~\ref{fig:residual_Re_dipole_frozen_core_CBS_def2tqz_av45z} and the RMSEs reported in Fig.~\ref{fig:RMSE_class_frozencore_CBS_def2tqz_av45z_dipole_Re}, frozen-core CCSD(T)/extrapolated-CBS results are not accurate for equilibrium distances of alkali-metal dimers, halogen-alkali metal and some transition metal diatomics. The complications to describe alkali-metal dimers can be rationalised by their weak bonding that requires a very accurate treatment of the electronic correlation. Transition metal diatomic outliers are expected, arising primarily due to inadequate treatment of multi-reference states. These results indicate that frozen-core CCSD(T)/extrapolated-CBS should not be used as a theoretical standard for geometries of weakly bound systems. Similarly, it should be used with caution for highly ionic systems, transition-metal-containing and multi-reference systems.



On average, the predicted frozen-core-CCSD(T)/CBS(def2-T/QZVPP) dipole moments (as computed at the calculated equilibrium distances) show an RMSE of 0.60~D, with errors commonly below 0.2 D (non-metal diatomics, metal/metalloid-nonmetal molecules, and some transition metal halides in def2- results). However, alkali-dimers, alkali-earth-metal-containing species, and transition-metal-containing diatomics display more significant errors. Thus, confirming the absence of correlation between experimental dipole moment and residual error. Similarly, we notice that dipole moments are generally far poorly predicted than equilibrium distances.

\begin{figure}[h!]
    \centering    
    \includegraphics[width=0.5\textwidth]{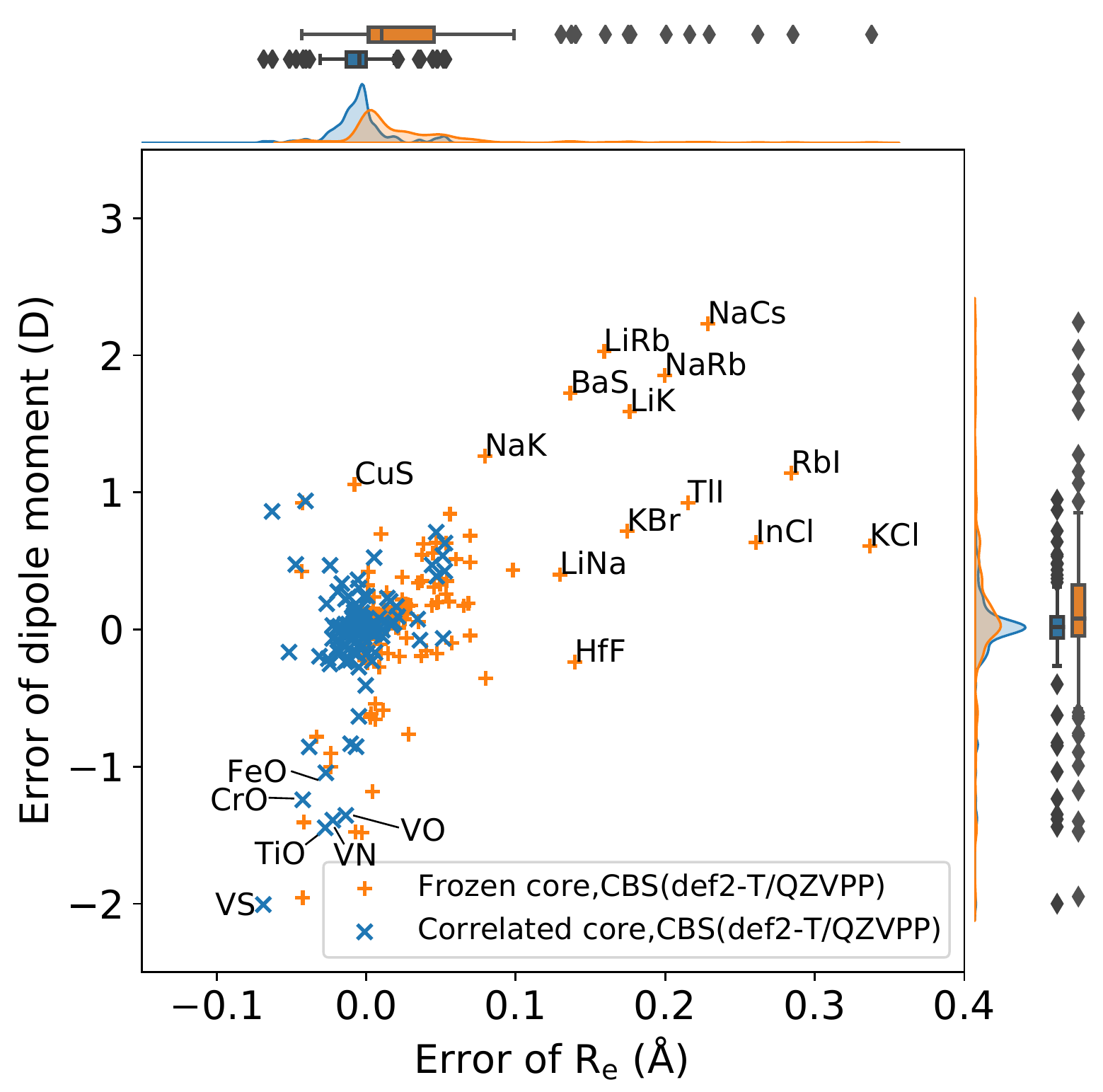}
    \caption{Errors of CCSD(T)/CBS(def2-T/QZVPP) dipole moments as a function of errors of computational optimized $R_e$ for 135 molecules, with or without core-correlations.}
	\label{fig:error_dipole_Re}
\end{figure}

The errors in equilibrium distance and dipole moment are not necessarily correlated, as shown in Fig.~\ref{fig:error_dipole_Re}. For instance, transition metal-non metal diatomics have relatively small RMSE for $R_e$ (0.03 \AA{}) but by far the largest RMSE for dipole moment (1.14 D), which may be related to the dominant covalent nature of the molecular bond in such systems. This result is significant because it suggests that benchmarking studies based only on energetic properties can fail to predict performance for other electron density properties.

We find that Dunning-extrapolated and def2-extrapolated results are generally quite similar (overall RMSE of 0.025 and 0.026 \AA{} for $R_e$, respectively) since both sets of results are expected to be close to the complete basis set limit. There are a few exceptions: CuS, NaF and NaCl, implying basis set incompleteness may still be an issue for some systems.


\begin{figure*}[h!]
    \centering     
    \includegraphics[width=1\textwidth]{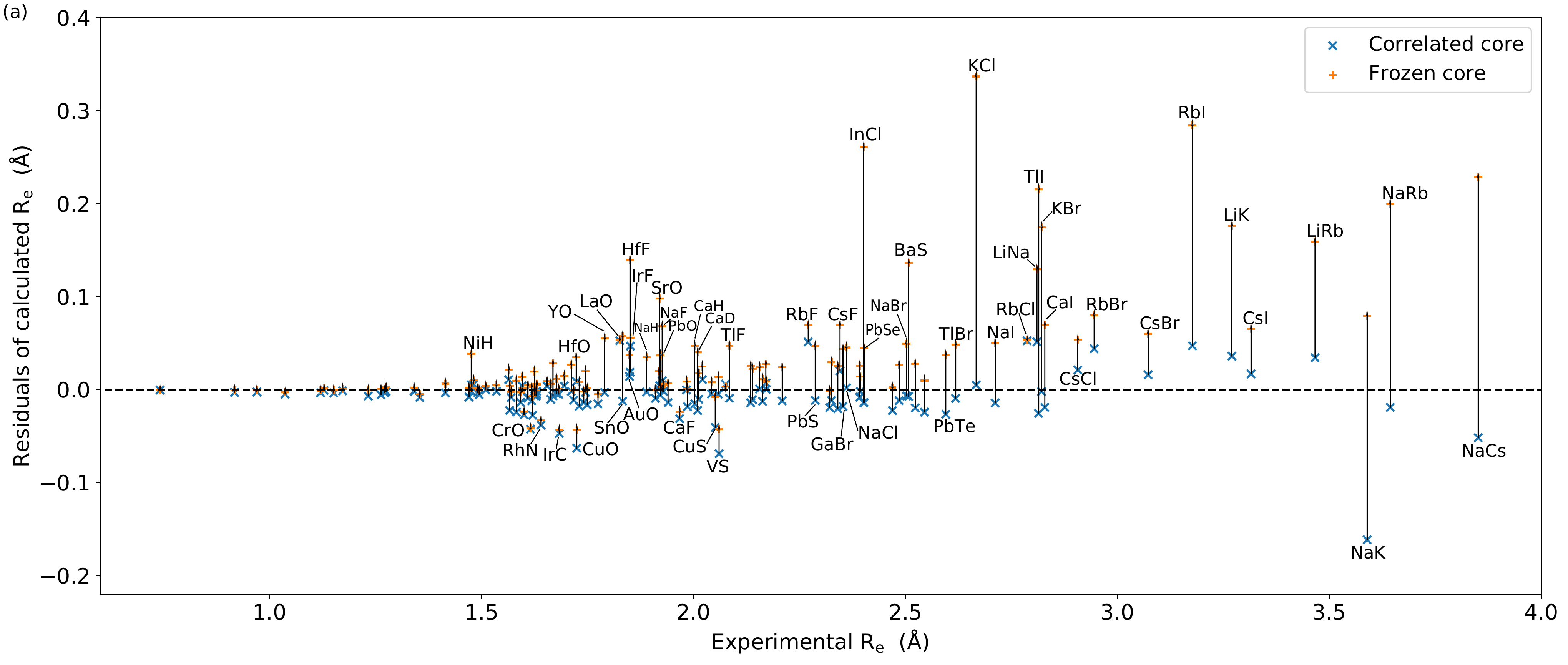}\\ 
    \includegraphics[width=1\textwidth]{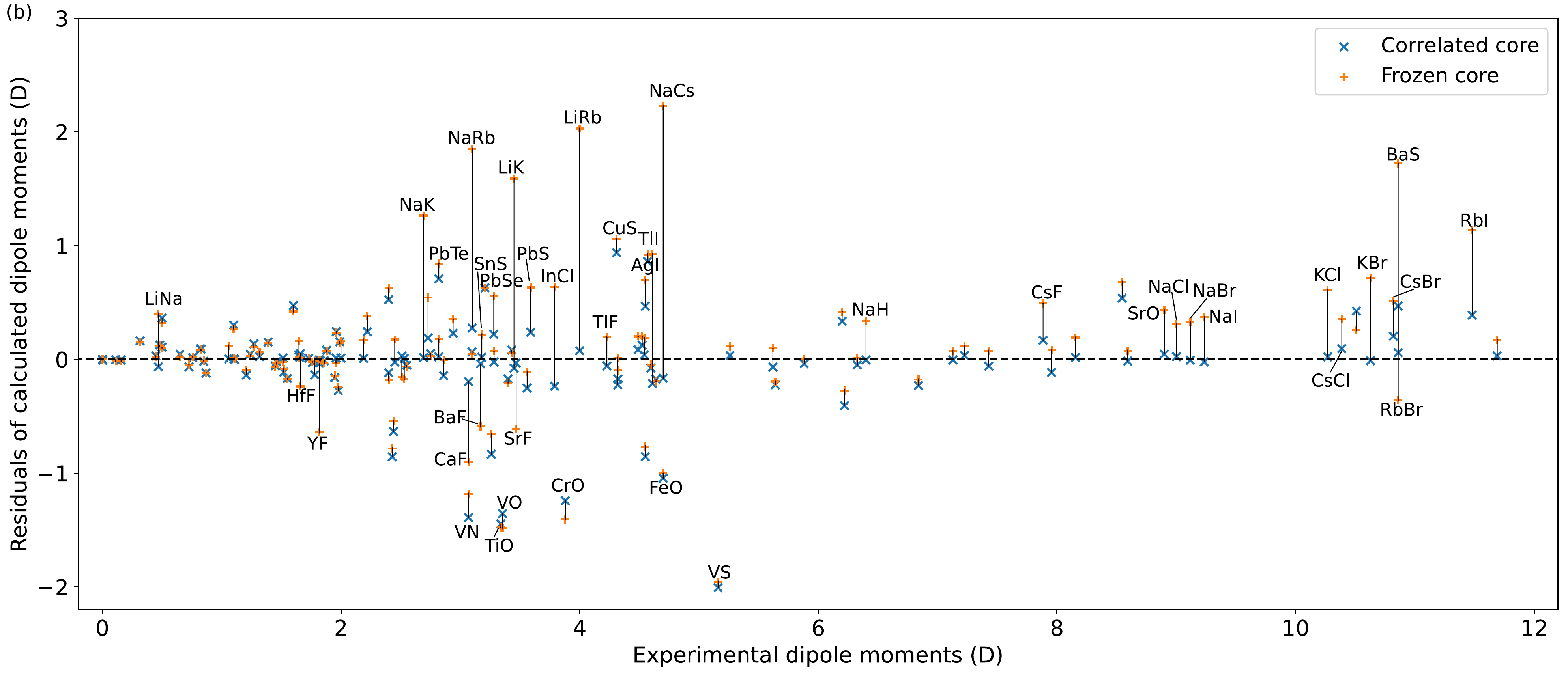}
    \caption{Residuals of calculated (a) $R_e$ and (b) dipole moments for 135 molecules by the predicted CBS(def2-T/QZVPP) with or without core-correlations. To guide the eyes, vertical lines are added to connect the dipole moments for the same molecules with or without core-correlations.}
	\label{fig:residual_dipole_core_CBS_def2tqz}
\end{figure*}

\subsection{\label{sec:core0}The role of core-correlations}

Core-core and core-valence correlation effects can be essential to molecules involving elements beyond the first row. However, frozen-core approximation are widely applied, especially for the systems involving heavy elements. For example, for the molecules with elements Z$>36$, ECPs are employed in combination with the def2- basis. However, for these molecules still, the inner valence shells are usually frozen in the calculations. Therefore, it is vital to quantify the influence of core-correlations on dipole moment and equilibrium distance computations.

\begin{figure}[h!]
    \centering    
    \includegraphics[width=0.5\textwidth]{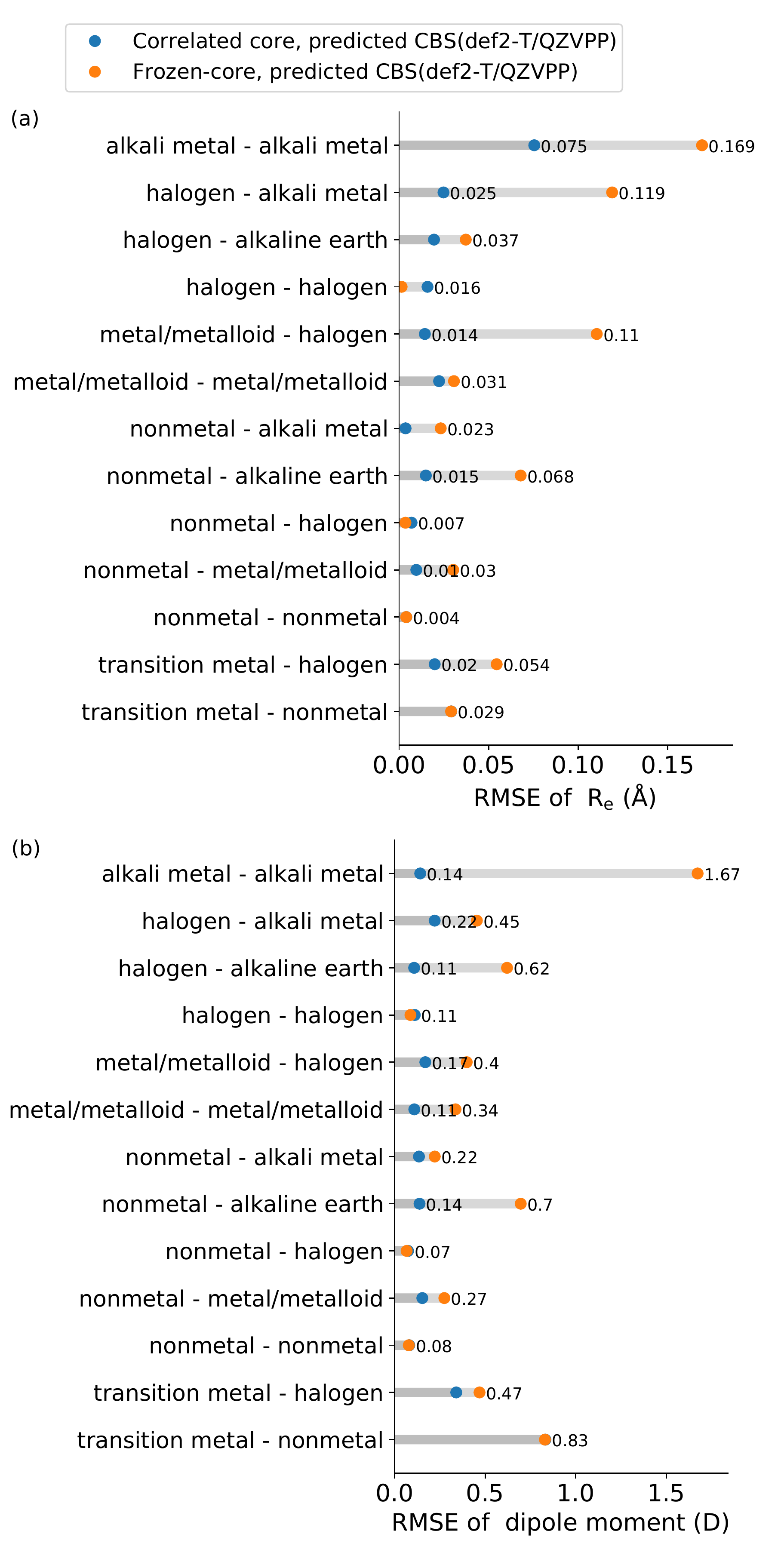}
    \caption{Comparison between RMSE of calculated (a) $R_e$ and (b) dipole moments by the predicted CBS(def2-T/QZVPP) with or without core-correlations for 135 molecules, classified by the types of molecules.}
	\label{fig:RMSE_class_core_CBS_def2tqz}
\end{figure}

Regarding the equilibrium distance, we observe a more accurate results in comparison with the experimental data when the core-correlation is taken into account, as shown in panel (a) of Fig.~\ref{fig:residual_dipole_core_CBS_def2tqz} and Fig.~\ref{fig:RMSE_class_core_CBS_def2tqz}. Core-correlations effects translate into a better prediction for most alkali metal halides, bialkali molecules, and metal/metalloid halides. Similarly, the predictions for alkaline earth compounds are also much better when including core-correlation effects. On the contrary, for transition metal-non metal compounds, the influence of core-correlations is minimal.

Fig.~\ref{fig:residual_dipole_core_CBS_def2tqz} compares the predicted CBS(def2-T/QZVPP) of dipole moments for the 135 molecules in the dataset. Interestingly enough, after including the core-correlations, we observe a dramatic improvement for bi-alkali ($\sim 1.5$D), alkaline halides ($\sim 0.2$D) and alkaline earth compounds ($\sim 0.5$D). In the case of bi-alkali molecules the heavier the metal atom is the better the improvement is. For halides, the molecules showing a larger ionic character are the ones that are better described. Therefore, an accurate description of the core-electron density can be essential for the accuracy of valence electron density.

For molecules involving elements with $Z>36$, we observe an improvement over frozen-core dipole moments, for example in HfF, YF, SrF, PbSe, PbS, TlI, TlF, AgI, InCl, even though the optimization space of correlated-core CCSD(T) wavefunctions is limited by the employment of ECPs. On the contrary, for the lighter 3d transition metal compounds, the errors of dipole moments can hardly be healed by including core-correlations, in agreement with our findings regarding the equilibrium distance.

Overall, our findings suggest that CCSD(T) core-correlations are essential in weakly-bound and highly ionic molecules for reliable predictions of geometries and dipole moments. For systems involving elements with $Z>36$, especially main-group compounds, taking into account core-correlations can still be very helpful. However, for most transition metal compounds, especially the 3d transition metal compounds, one can hardly benefit from including core-correlations.

\begin{figure*}[h!]
    \centering     
    \includegraphics[width=1\textwidth]{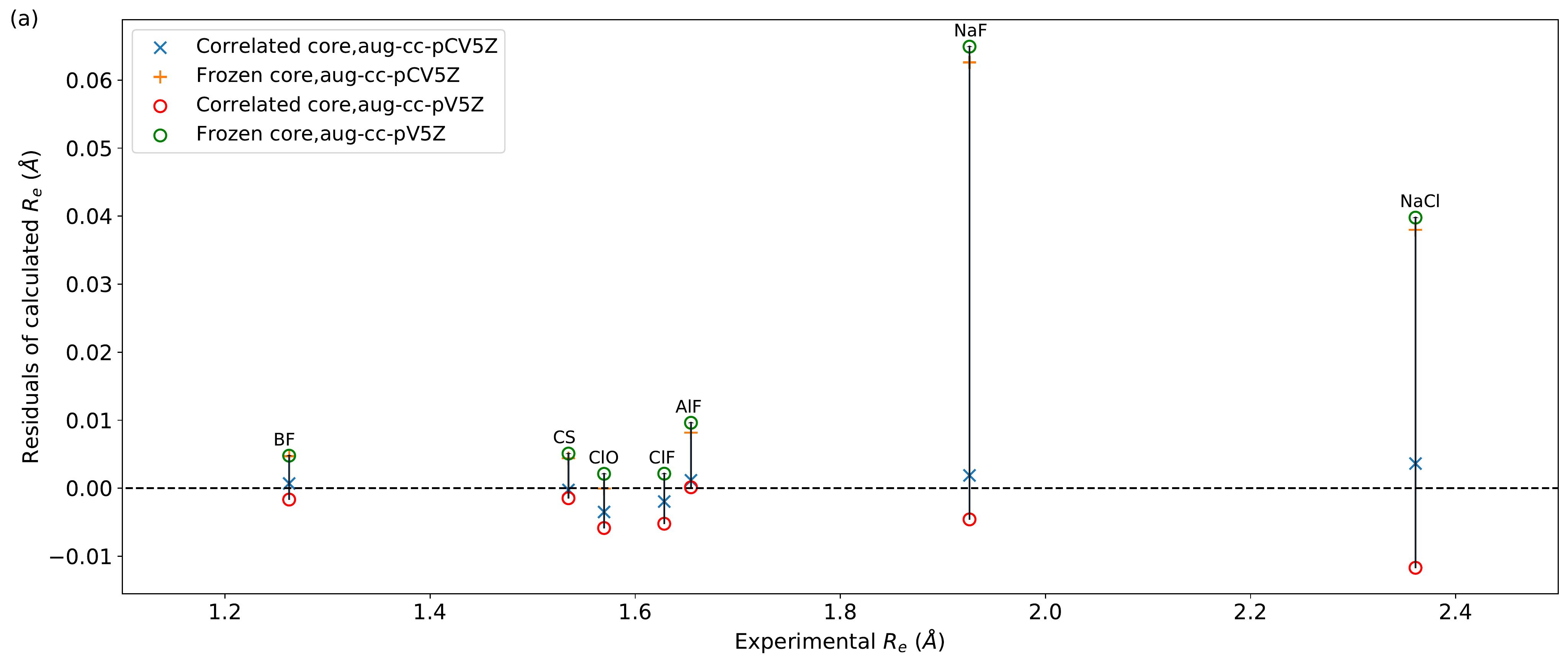}\\ 
    \includegraphics[width=1\textwidth]{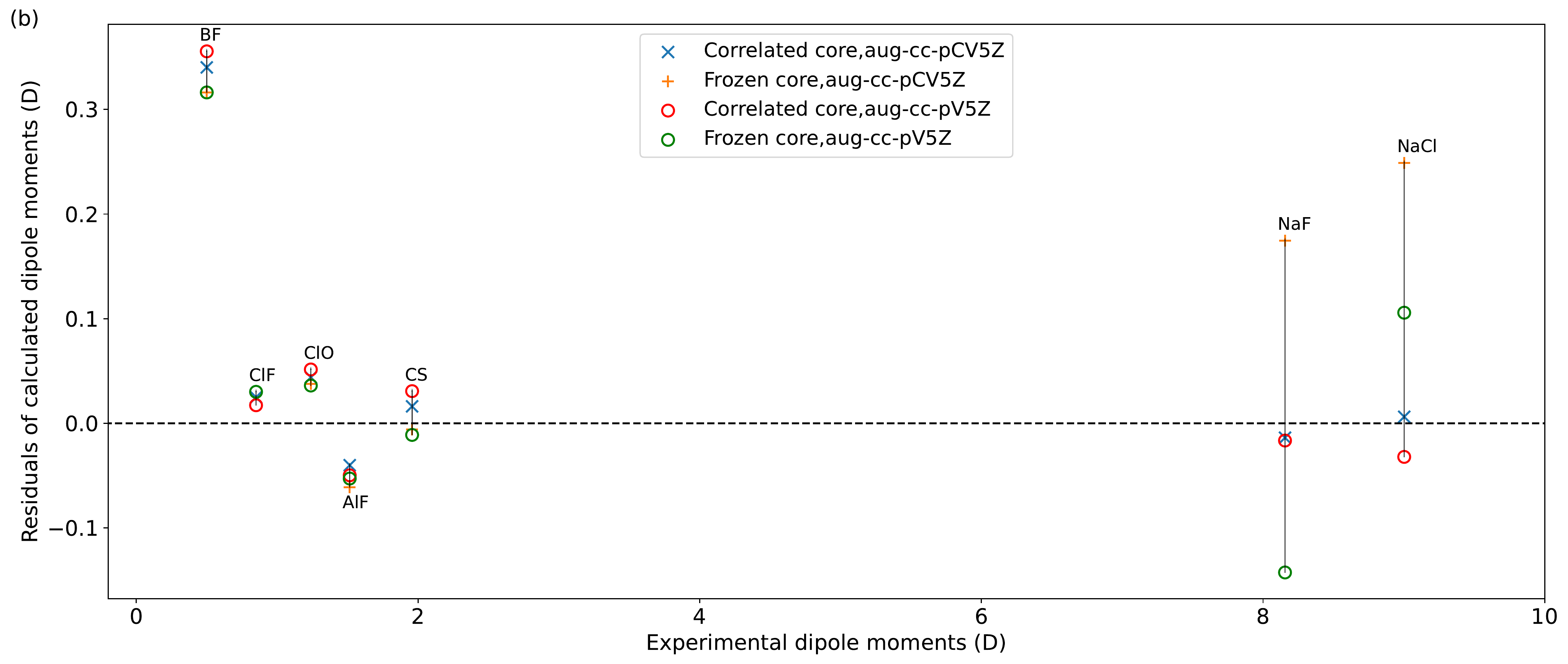}
    \caption{Residuals of calculated (a) $R_e$ and (b) dipole moments for 8 molecules with aug-cc-pV5Z and aug-cc-pCV5Z basis sets, with or without core-correlations. To guide the eyes, vertical lines are added to connect the dipole moments for the same molecules with or without core-correlations.}
	\label{fig:residual_dipole_Re_augccpcv}
\end{figure*}

To further investigate the influence of core-correlations, we use the aug-cc-pCV basis to calculate $R_e$ and dipole moments of 7 molecules consisting of second and third-row main-group elements, shown in Fig.~\ref{fig:residual_dipole_Re_augccpcv}. Compared to aug-cc-pV basis, additional basis functions are included in aug-cc-pCV basis optimized for core-core and core-valence correlations so that all the electrons not replaced by ECP are correlated. As expected, the influence of the additional correlating functions for the core electrons to $R_e$ is small under frozen-core approximation, whereas the difference of $R_e$ with aug-cc-pCV5Z and aug-cc-pV5Z becomes larger when core-correlations are taken into account in the calculations. However, for dipole moment, as it is shown in Fig.~\ref{fig:residual_dipole_Re_augccpcv} (b), we observe slight improvement from the additional core-valence functions for most of the seven molecules except for NaF and NaCl. Actually, under the frozen-core approximation, the dipole moments of NaF and NaCl predicted by aug-cc-pCV5Z, and aug-cc-PV5Z basis sets are very different. Although, when we consider the core-correlations, the dipole moments predicted by the two basis converge to be very closed to the experimental values.

\subsection{\label{sec:cbs}The basis set incompleteness error}



The Dunning basis sets are expected to converge to the CBS regarding energetic properties systemically. However, the convergence behavior of the def2- series basis sets is not fully documented, especially for dipole moments of diatomics involving 4th-6th row elements, including transition metal elements. In the meantime, it has been shown that for some light-element molecules, using two-point extrapolations from def2-T/QZVPP to the CBS, the accuracy of def2- predictions can be improved for the self-consistent field energies \cite{neese2011revisiting}. However, for dipole moment, which depends on the properties of electron densities, it is not yet clear if one can benefit from such extrapolations.





\begin{figure}[h!]
    \centering    
    \includegraphics[width=0.43\textwidth]{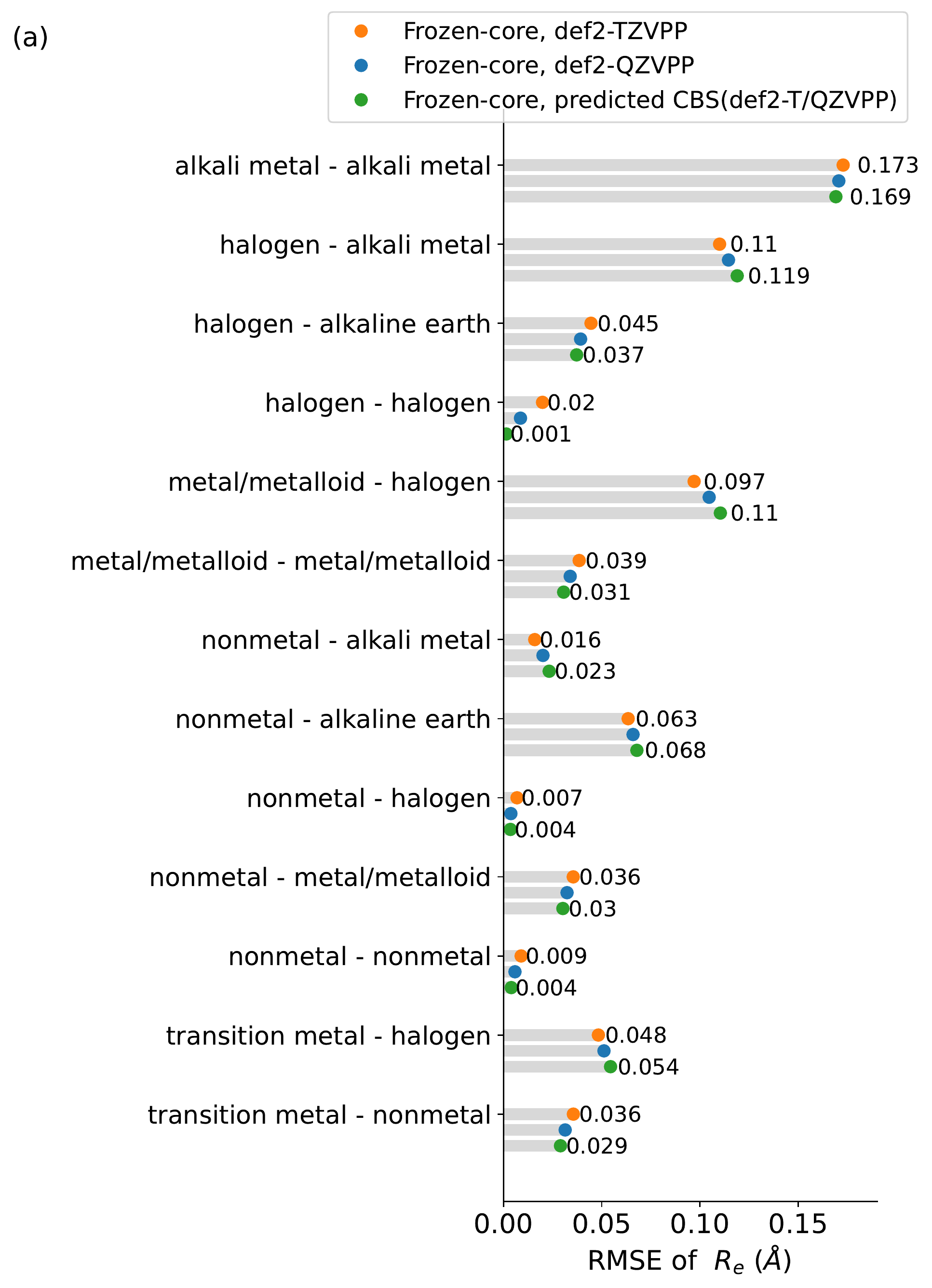}  
    \includegraphics[width=0.43\textwidth]{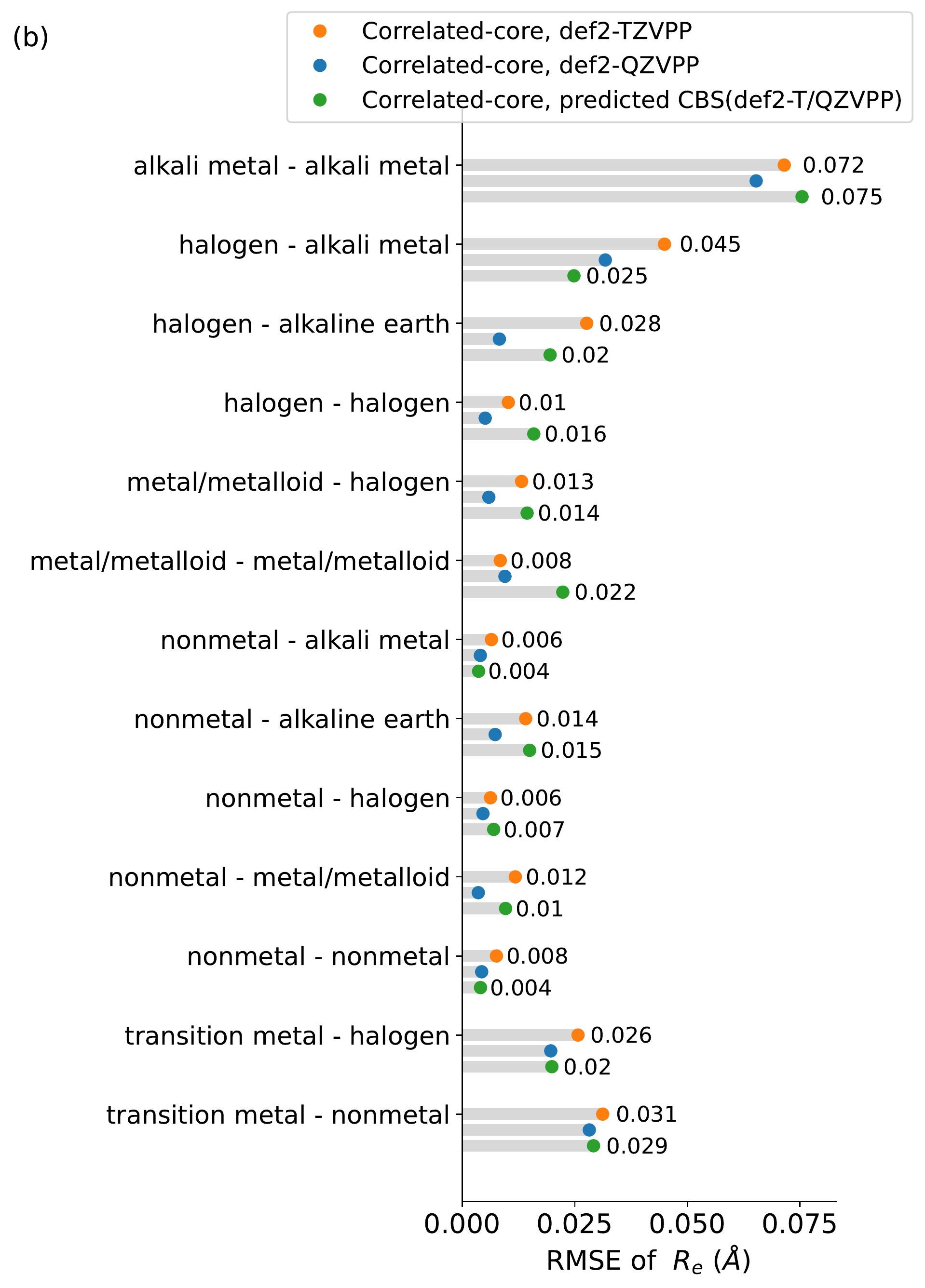}
    \caption{Comparison between RMSE of calculated $R_e$ obtained from different levels of def2- basis set (a) under frozen-core approximation or (b) with core-correlations.}
	\label{fig:CBS_convergence_Re_class_def2tqz}
\end{figure}

As discussed above, we get the predicted CBS' from the inverse cubic extrapolation. As shown in panel (a) of Fig.~\ref{fig:CBS_convergence_Re_class_def2tqz}, under frozen-core approximation, the equilibrium distances $R_e$ are almost converge at the def2-TZVPP level, where the basis set incompleteness error is smaller than $0.005$ \AA{}, in agreement with common observations in quantum chemistry calculations. Nevertheless, 8 out of the 13 classes of molecules can be slightly improved by extrapolating to the predicted CBS(def2-T/QZVPP), especially halogen diatomics. However, in main-group and transition metal halides, the predicted $R_e$ at the def2-QZVPP level are worse than the def2-TZVPP results, leading to larger errors when the CBS(def2-T/QZVPP) is employed. Therefore, one can hardly benefit from the CBS extrapolation in these cases.

When core-correlation is included, as shown in panel (b) of Fig.~\ref{fig:CBS_convergence_Re_class_def2tqz}, the convergence behavior of $R_e$ becomes more complicated than the frozen-core cases. In most cases, $R_e$ obtained from def2-QZVPP calculations are the closest to experimental values, and improvements by using def2-QZVPP instead of def2-TZVPP basis are significant. However, the predicted CBS(def2-T/QZVPP) generally gives larger errors than def2-QZVPP, suggesting the breakdown of the cubic extrapolation scheme. There are a few exceptions, including alkali metal halides, for which the errors of $R_e$ can be reduced for 0.02 \AA{} at the CBS compared to def2-TZVPP level, while for alkali metal-nonmetal compounds, the improvements are 10 times smaller. Therefore, for core-correlated calculations of $R_e$, def2-QZVPP is the optimal choice for geometry optimization.

For dipole moments, the convergence behavior becomes very different than for $R_e$. Under frozen-core approximation, the overall RMSE can be slightly reduced from 0.63 to 0.60 D using def2-TZVPP and def2-QZVPP basis, respectively. However, in most main-group diatomics expect alkali metal halides, one can hardly benefit from using the larger def2-QZVPP basis nor the CBS extrapolation. Especially for alkaline earth-nonmetal compounds, the predicted CBS(def2-T/QZVPP) errors are 0.25 D larger than def2-TZVPP dipole moments. On the contrary, for molecules involving transition-metal elements, def2-QZVPP generally gives better results can def2-TZVPP, and the predicted CBS(def2-T/QZVPP) gives the best predictions. Therefore, it would be better for most main-group molecules to stop at the def2-TZVPP level, while for transition-metal compounds, def2-QZVPP can give better results. However, when calculating dipole moments with core-correlations, improvements have been observed by using the def2-QZVPP basis over def2-TZVPP basis. The most significant improvements occur in bi-alkali, alkali metal halides and alkaline earth halides, where the errors can be reduced for around 0.15 D. In the meantime, the cubic extrapolation to CBS gives the best predictions of dipole moments in most cases except some halogen diatomics, nonmetal halides and alkaline earth halides. 

In summary, as a property of electron density, the dipole moment is more sensitive to the size of basis sets than the molecular geometry. When considering core-correlations, both $R_e$ and dipole moments can be improved when increasing the size of the basis set from def2-TZVPP to def2-QZVPP. Therefore, the CBS extrapolation is especially suggested for correlated-core dipole moments of alkali metal compounds, while for systems involving transition metals, the extrapolated CBS is the optimal choice either with or without core-correlations.

\begin{figure}[h!]
    \centering    
    \includegraphics[width=0.43\textwidth]{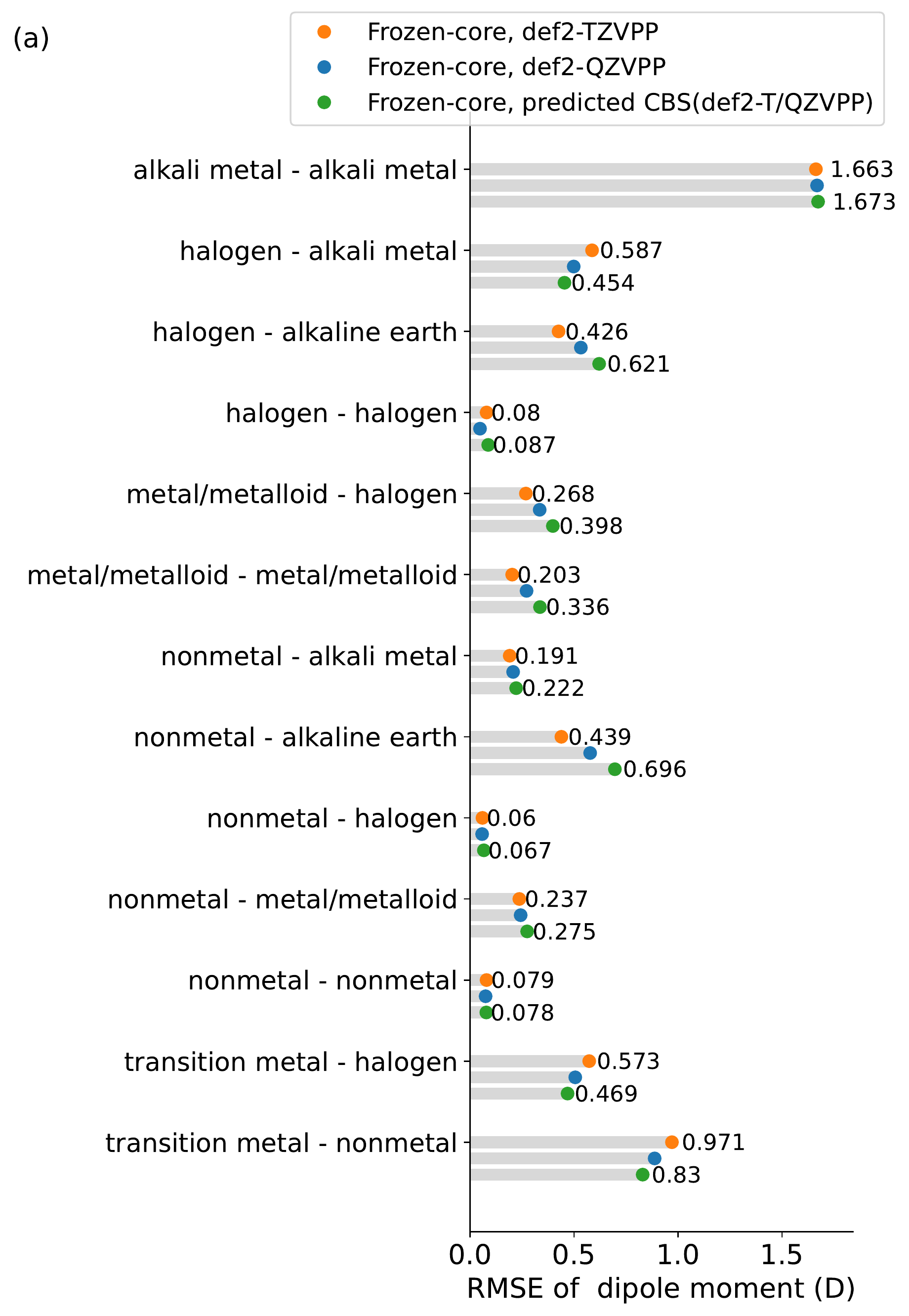}  
    \includegraphics[width=0.43\textwidth]{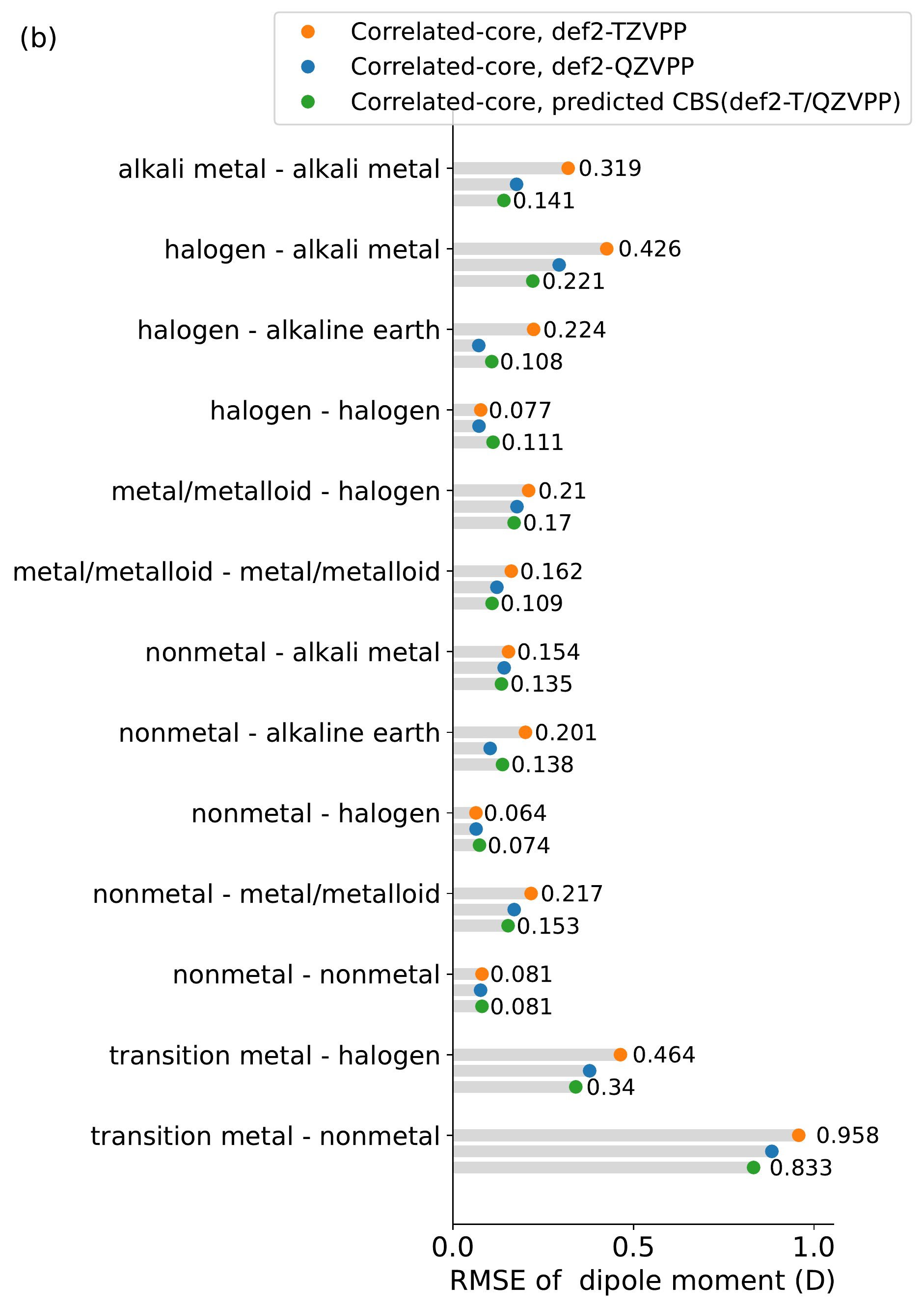}
    \caption{Comparison between RMSE of calculated dipole moments obtained from different levels of def2- basis set (a) under frozen-core approximation or (b) with core-correlations.}
	\label{fig:CBS_convergence_dipole_class_def2tqz}
\end{figure}

\subsection{\label{sec:diffuse}Influence of diffuse functions}

It has been demonstrated that for molecules consisted of first-three-row elements, the diffuse function augmentations to the Dunning basis can provide significant improvements to electron-density-related properties, including dipole moments \cite{peterson1997co, hickey2014benchmarking, zapata2020computation}, as they decay slowly with increasing internuclear distance. For example, for CO, adding extra diffuse augmentations improves the systematical convergence of its CCSD(T) $R_e$ and dipole moment, although the improvements are already negligible at the QZ level \cite{peterson1997co}. 

For the present dataset, as discussed above, the predicted CBS(def2-T/QZVPP) overestimates both $R_e$ and dipole moments of bi-alkaline molecules. Considering the van der Waals nature of these molecules, one might expect that such overestimation can be healed by introducing diffuse augmentations into the basis sets better to describe the long-range tails of the wave functions. To test this idea, we examine the performance of def2-QZVPPD basis sets\cite{def2qzvppdrappoport2010property} in 16 selected molecules, 5 of which are bi-alkaline molecules. As a result, and surprisingly enough, as shown in Fig.~S3 and Table S1, the use of diffuse functions in the def2-QZVPPD basis sets does not improve $R_e$ nor dipole moments in our tested cases, even if core-correlations are included. 

\begin{figure*}[h!]
    \centering    
    \includegraphics[width=1\textwidth]{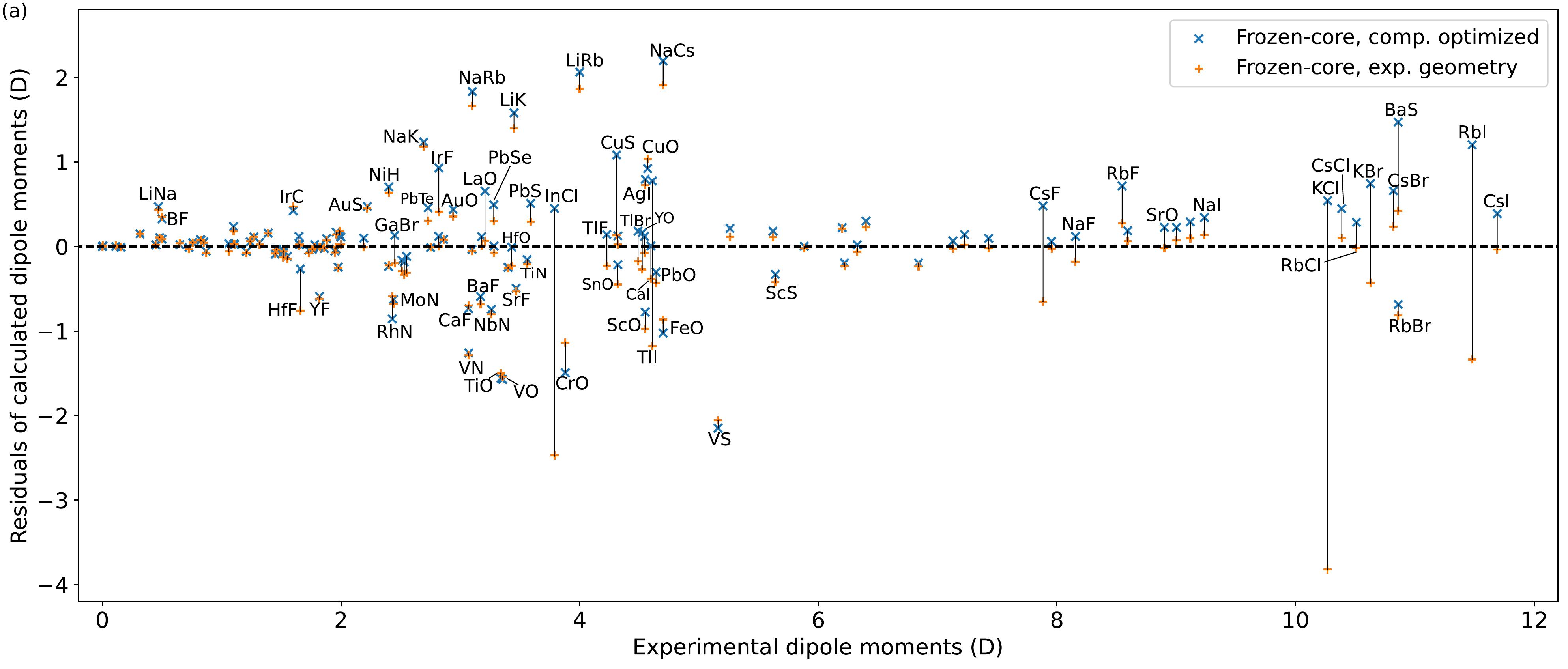}\\  
    \includegraphics[width=1\textwidth]{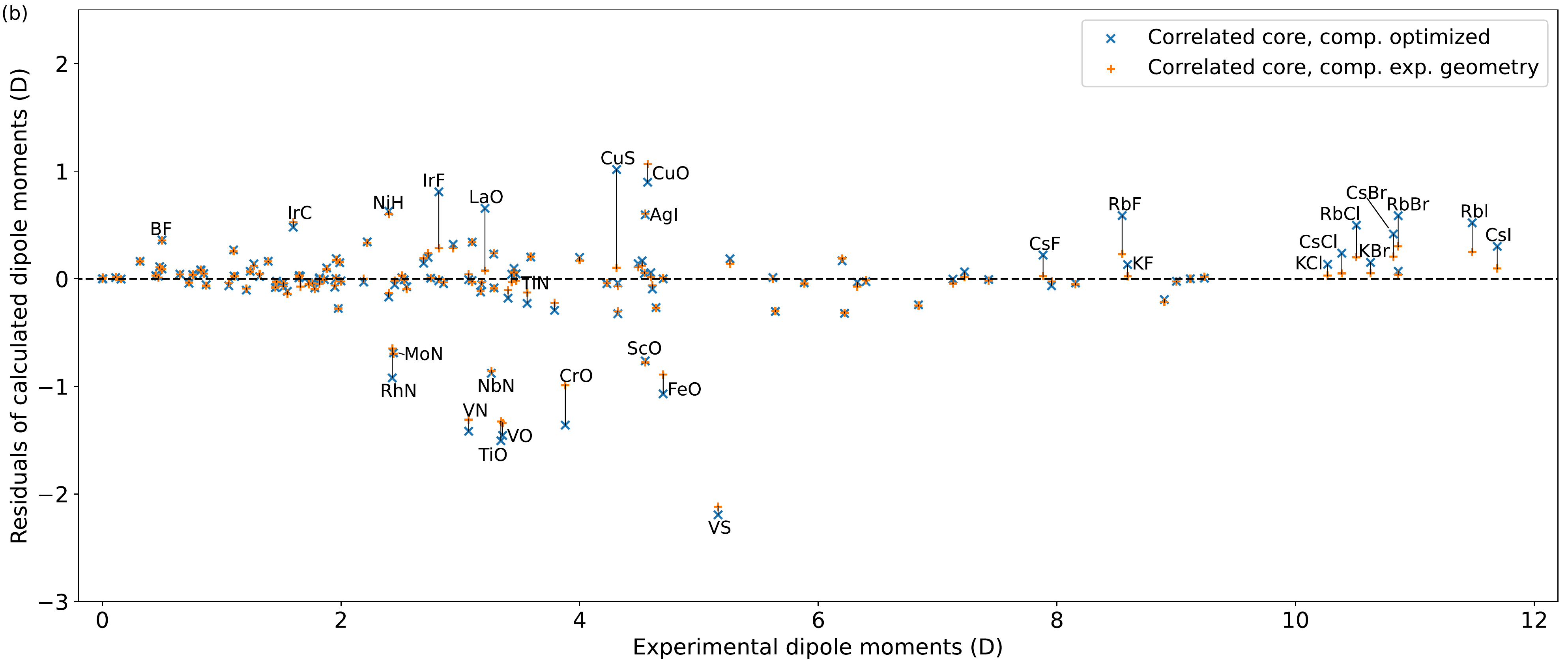}
    \caption{Residuals of calculated dipole moments for 135 molecules by def2-QZVPP basis set with computational optimized or experimental geometries (a) under frozen-core approximation or (b) with core-correlations. To guide the eyes, vertical lines are added to connect the dipole moments for the same molecules.}
	\label{fig:residual_dipole_frozencore_core_exp_vs_opt_def2qzvpp}
\end{figure*}

\subsection{\label{sec:opt_vs_exp_Re}Dipole moments from experimental geometries}

\begin{figure}[h!]
    \centering    
    \includegraphics[width=0.5\textwidth]{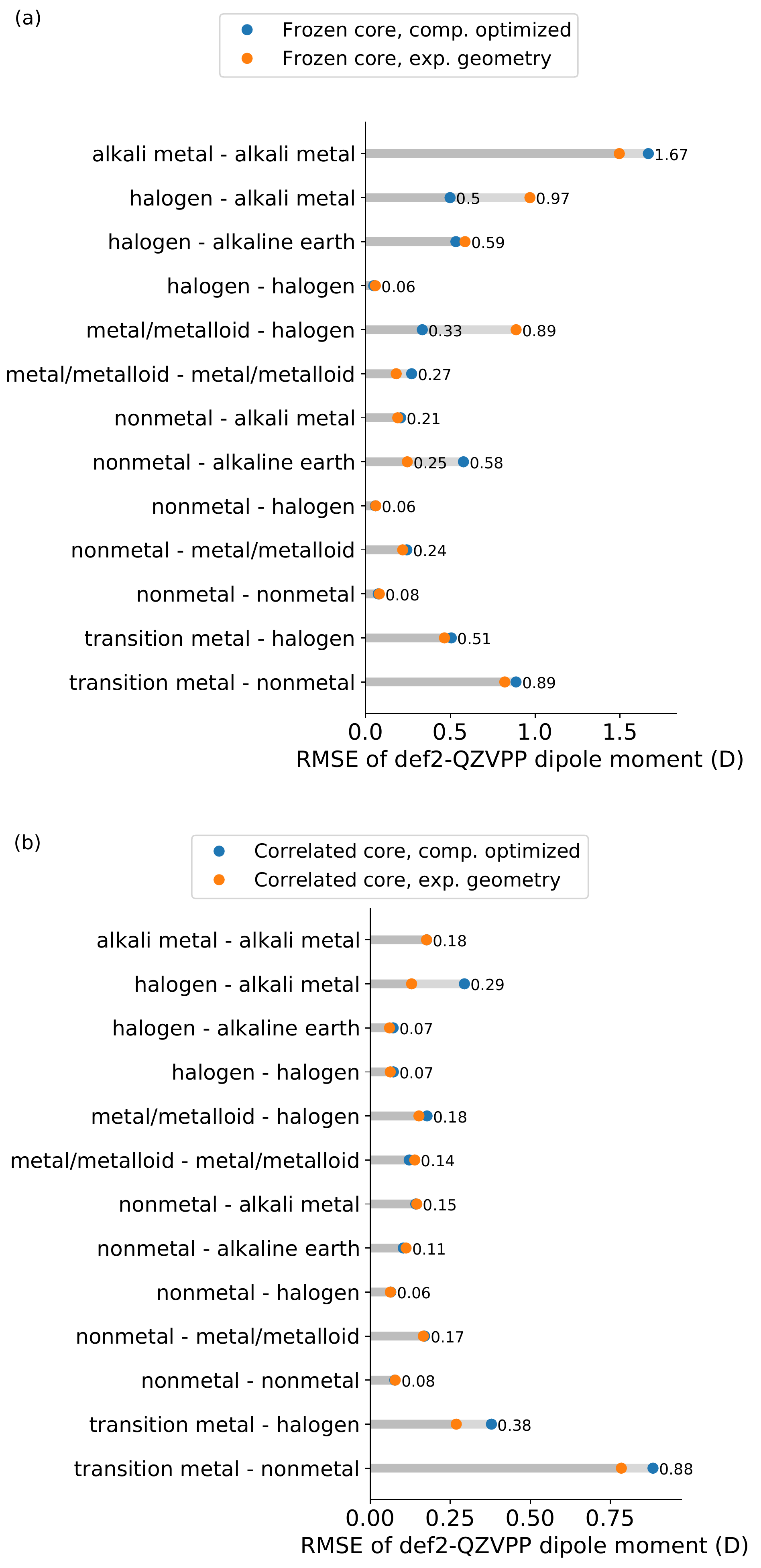}
    \caption{Comparison between RMSE of calculated dipole moments obtained from computational optimized or experimental geometry (a) under frozen-core approximation or (b) with core-correlations, with the def2-QZVPP basis set.}
	\label{fig:RMSE_class_frozencore_core_opt_vs_exp_Re_def2qzvpp}
\end{figure}

The dipole moments discussed in previous sections are obtained from optimized computational geometries at the corresponding levels of theory and basis sets. Nevertheless, and despite our exhaustive work, we did not analyze if the performance of the different methods and basis set for the dipole moment were correlated with their performance for the equilibrium geometry. For instance, when we consider the core-correlation, both magnitudes showed a simultaneous improvement. To reach a deeper insight on that question, we have calculated the dipole moment of the molecules in the dataset using experimental geometries taken from Ref.~\cite{dipolePCCP2020} and compare with the ones based on computational optimized geometries, as shown in Figs.~\ref{fig:residual_dipole_frozencore_core_exp_vs_opt_def2qzvpp} and Fig.~\ref{fig:RMSE_class_frozencore_core_opt_vs_exp_Re_def2qzvpp}. For these calculations, we employ the def2-QZVPP basis since it has been shown in the previous sections that the def2-QZVPP dipole moments are very closed to the predicted CBS(def2-T/QZVPP).

Within the frozen-core approximation, the calculated dipole moments based on experimental geometries are generally smaller than the ones employing computational optimized geometries. Consequently, using experimental geometries can improve the dipole moments of some molecules whose dipole moments are overestimated, for example CuS, some bi-alkali molecules and some of the alkaline halides, but exacerbates the description of some metal/nonmetal halides. However, the RMSE is 0.67 D, which is even larger than the dipole moments for computational optimized geometries (0.61 D). In the case of core-correlations, as the description of dipole moments is already improved compared to the frozen-core results, using experimental geometries can make the predictions of some molecules closer to the experimental results, for example some transition metal compounds whose dipole moments are overestimated with computational optimized geometries. The overall RMSE of dipole moment can be reduced from 0.42 D with computationally optimized geometries to 0.37 D with experimental geometries for 135 molecules.

\section{Summary and Conclusion}

In this study, we have systematically analyzed the performance of CCSD(T) predicting dipole moments and equilibrium distances of diatomic molecules compared with a dataset of 135 experimental dipole moments and equilibrium distances. In particular, we have employed the def2- basis to explore the accuracy at the predicted CBS limits. Similarly, we have tested the predictions of the aug-cc-pV basis for 67 molecules containing elements between the first and fourth row of the periodic table. As a result, we find that CCSD(T) fails to assess some diatomic molecules' dipole moment under frozen-core approximation, in particular transition metal-nonmetal compounds, bi-alkali molecules, and main-group metal halides. Indeed, we find that these errors can hardly be reduced by using experimental geometries. For bi-alkaline and main-group metal halides, it is mandatory to include the core-correlation to reach a proper agreement with experimental determined dipole moments (with errors $<$ 0.25 D) and equilibrium distances. However, even if the core-correlations are considered, the errors predicting dipole moments of transition metal compounds are still large. Similarly, surprisingly enough, after including augmented diffuse functions, we did not observe any improvement of the predicted magnitude versus the experimental results.

CCSD(T) is the primary benchmarking method of choice to study the performance of different quantum chemistry tools such as DFT or machine learning-based methods. However, our results indicate that CCSD(T) quality as a benchmark depends very much on the molecule and approximations under consideration. In particular, the default fronzen-core CCSD(T) approach does not lead to satisfactory results for dipole moments. Therefore, it is of prime interest to analyze the nature of the system under consideration before using CCSD(T) as a benchmark. In the same vein, we conclude that to have an adequate description of the dipole moment in diatomics it is necessary to include core-correlations effects and use a basis set explicitly accounting for the core electrons, which makes the calculations more cumbersome than expected.

\begin{acknowledgement}

The authors thank  Dr. Adam Wasserman and Yuming Shi for their insight and suggestions on this work, and Dr. Stefan Truppe for fruitful discussions. Finally, X.L and J.P-R. acknowledge Prof. Dr. Gerard Meijer for supporting our work.

\end{acknowledgement}


\begin{suppinfo}
\section*{Data availability}

The data underlying this article are available in the article and in its online supplementary material. These include the following files;
\begin{itemize}
    \item Dipole moments from $ab initio$ calculations\\ (data\_dipole\_moment\_CCSDt\_135mols.csv);
    \item Optimized equilibrium internuclear distances\\ (data\_Re\_CCSDt\_135mols.csv)
    \item \textit{Ab initio} input and output files with the def2-QZVPP basis\\ (original\_data\_def2qzvpp.zip)).
\end{itemize}

\section*{Additional Supporting Information}
As well as the data files described above, we include as supporting information five additional figures, three additional tables and associated discussions.

\end{suppinfo}

\bibliography{achemso-demo}

\end{document}